\documentclass{article}

\usepackage{arxiv}

\usepackage[utf8]{inputenc} 
\usepackage[T1]{fontenc}    
\usepackage{hyperref}       
\usepackage{url}            
\usepackage{booktabs}       
\usepackage{amsfonts}       
\usepackage{nicefrac}       
\usepackage{microtype}      
\usepackage{lipsum}
\usepackage{graphicx}
\graphicspath{ {./images/} }
\usepackage{natbib}
\usepackage{amsmath}
\usepackage{float}
\usepackage{multirow}

\title{Green Finance and Carbon Emissions: A Nonlinear and Interaction Analysis Using Bayesian Additive Regression Trees}

\author{
  Mengxiang Zhu \\
  School of Mathematics and Statistics \\
  University College Dublin \\
  Belfield, Dublin 4, Ireland, D04V1W8 \\
  \texttt{mengxiang.zhu@ucdconnect.ie}
  \and
  Riccardo Rastelli \\
  School of Mathematics and Statistics \\
  University College Dublin \\
  Belfield, Dublin 4, Ireland, D04V1W8 \\
  \texttt{riccardo.rastelli@ucd.ie}
}

\begin{document}
\maketitle
\begin{abstract}
This study investigates the nonlinear and interaction effects of green finance on carbon emission intensity (CEI) using Chinese provincial panel data from 2000 to 2022. The Climate Physical Risk Index (CPRI) is incorporated into the analytical framework to assess the potential influence of climate risk on carbon emission performance. We employ Bayesian Additive Regression Trees (BART) to capture complex nonlinear relationships and interaction pathways between the variables, then use SHapley Additive exPlanations to enhance model interpretability. Results show that the Green Finance Index (GFI) has a significant effect on CEI, exhibiting an inverted U-shaped effect, with notable regional heterogeneity. Contrary to expectations, there is insufficient evidence to suggest a significant association between CPRI and CEI. Further analysis reveals that in high energy consumption scenarios, stronger green finance development contributes to lower CEI. These findings highlight the potential of green finance as an effective instrument for carbon reduction, especially in energy-intensive contexts, and underscore the importance of accounting for nonlinear effects and regional disparities when designing and implementing green financial policies.

\end{abstract}

\section{Introduction}

With the escalation of global climate change and the increasing frequency of extreme weather events, reducing carbon emissions has emerged as an urgent global priority. Green finance, as a vital strategic instrument in addressing climate change, seeks to channel financial resources into environmentally sustainable projects aimed at improving energy efficiency and reducing carbon emissions \cite{wang2023can,wu2024impact}. Green finance encompasses a range of financial instruments, including green credit, green bonds, and sustainable investment funds, which effectively allocate capital to projects that mitigate the adverse impacts of climate change \cite{niyazbekova2021growth}.
As the world’s largest carbon emitter, China has positioned green finance as a key driver in its transition toward a low-carbon economy, particularly in pursuit of its dual carbon targets: achieving carbon peaking by 2030 and carbon neutrality by 2060 \cite{lin2023china,du2025decarbonization}. According to the report \cite{Institute2023Fintech}, China has become the world’s largest green credit market and the second-largest green bond market, with the scale of green finance continuing to grow. However, the actual effectiveness of green finance is strongly shaped by local economic structures and environmental conditions. Recent studies suggest that climate risks significantly increase market uncertainty \cite{lee2024impact} and influence investor decision-making \cite{lu2025climate}, thereby affecting the carbon reduction effects of green finance. Nevertheless, climate risk factors are rarely incorporated into analytical frameworks assessing the impact of green finance on carbon emissions. Moreover, most existing studies rely on traditional linear econometric models with manually specified interaction terms, and few have systematically examined the potential nonlinear effects or interaction mechanisms of green finance on carbon emissions.

To address these research gaps, this study employs Bayesian Additive Regression Trees (BART), which is particularly suited to capture nonlinear relationships and complex variable interactions. Drawing on panel data covering multiple economic and environmental indicators across 30 Chinese provinces from 2000 to 2022, the analysis incorporates climate risk into the modeling framework to systematically examine how green finance affects carbon emissions and whether climate risk has a significant impact on emission outcomes.

This research makes three key contributions to bridge the gaps in variable integration and methodological innovation: first, it incorporates climate risk into the analytical framework to assess its direct effect on carbon emissions and its potential interaction effects with green finance; second, by employing BART, it overcomes the limitations of traditional linear econometric approaches in capturing nonlinear dynamics and interaction effects. Leveraging interpretable tools such as Partial Dependence Plots (PDPs) and SHapley Additive exPlanations (SHAP), this study reveals the underlying mechanisms through which green finance associates with carbon emissions and provides interpretable model-based evidence; third, methodologically, this study expands the analytical toolkit at the intersection of green finance and carbon emissions research, demonstrating the feasibility and value of machine learning approaches in environmental policy evaluation. As a consequence, this contribution offers technical guidance for future research and provides a robust empirical foundation for the formulation of green transition policies.

\section{Literature review}
\label{sec:literature review}
\subsection{Impact factors of carbon emissions}

Extensive research has been conducted to identify the driving forces of carbon emissions. Among the various analytical frameworks, the STIRPAT model of \cite{york2003stirpat} remains one of the most widely applied. \cite{yang2018what} found a temporal shift in key emission drivers from urbanization to foreign trade. \cite{tang2024study} further highlighted that in the metal smelting industry, population, coal consumption, and urbanization were the most sensitive factors. Through Granger causality tests, \cite{xiao2023decomposition} confirmed that economic growth, urbanization, industrial structure, and energy structure significantly influence carbon emissions. To better address spatial and temporal variations, \cite{chen2021analysis} adopted the GTWR-STIRPAT model and found that energy efficiency, trade, industrial structure, and environmental regulation exerted spatially heterogeneous and time-lagged effects. Similarly, \cite{li2022dynamic}, using the Spatial Durbin Model and Panel Threshold Model, demonstrated that energy consumption and non-green technologies drive short-term increases in carbon emissions, whereas green innovation and industrial upgrading play a key role in long-term emission reductions. In addition, several researchers have employed decomposition analysis to examine the structural components of carbon emissions. For examples, \cite{chen2019driving} identified several key driving factors, including potential carbon factor effects, energy intensity. \cite{liu2019driving} found that economic activity, emission efficiency, and potential carbon factors were strongly associated with emissions growth, while improvements in energy intensity, pure technical efficiency, and scale efficiency helped mitigate emissions. \cite{zhang2022carbon} concluded that economic efficiency is the sole driver of industrial carbon emissions, while energy intensity remains the most critical factor in carbon reduction. In contrast, \cite{sun2022characteristics} found that energy and industrial structures have relatively minor effects on the growth rate of carbon emissions, and that energy intensity is the primary factor contributing to decoupling.

\subsection{Green finance and carbon emissions}

Green finance, as an important financial tool to promote the low-carbon transition, has been widely confirmed by empirical studies to have a significant suppressive effect on carbon emissions. \cite{lin2023impact} found that green finance significantly reduces local carbon emissions and generates spillover effects in surrounding regions. Among various green financial instruments, \cite{wang2023can} argue that the carbon reduction effect of green credit is more pronounced than that of green bonds. Green finance not only contributes directly through investment in low-carbon projects, but also facilitates carbon reduction indirectly by stimulating green technological innovation, optimizing industrial structure, and improving energy efficiency \cite{wang2022how,tu2025impact,ran2023driving}.

However, the carbon reduction effects of green finance are not always consistent across different contexts. To explore its underlying mechanisms more deeply, recent studies have incorporated contextual variables into their analyses. \cite{zhao2024role} found that in regions with high political risk or high economic complexity, the effectiveness of green finance in reducing emissions is significantly weakened, suggesting that the institutional environment plays a moderating role. \cite{ma2024research} further emphasized that factors such as the level of governmental environmental attention and the development of the financial sector exhibit marked heterogeneity in shaping the effectiveness of green finance in influencing carbon emissions.

Moreover, researchers have begun to explore the nonlinear nature of green finance's impact. \cite{sun2022characteristics} and \cite{yu2024machine} found that the effect of green finance on carbon emissions is not a simple linear decline. Instead, it varies significantly under different conditions of energy consumption structure and financial technology development. Similarly, \cite{cai2025how} uncovered complex nonlinear interactions between green finance and other factors such as environmental investment, inclusive finance, and financial regulation, all of which significantly influence carbon emission intensity.

In summary, while the role of green finance in carbon emission reduction is increasingly evident, its underlying mechanisms exhibit notable complexity, regional heterogeneity, and nonlinearity.  

\subsection{Application of machine learning algorithms in carbon emission}

The existing literature has made significant progress in exploring the factors influencing carbon emissions using traditional econometric models, particularly the STIRPAT framework \cite{li2017industrial,zhang2019identifying,thio2022estimation}. However, these approaches generally rely on assumptions of linear relationships among variables and are often affected by researchers’ prior selection of explanatory factors. As a result, they have limitations in capturing complex nonlinear relationships and high-order interaction effects. In recent years, the development of machine learning methods has provided a novel technical pathway for modeling carbon emission mechanisms and identifying key drivers. Compared to traditional models, machine learning algorithms can automatically learn nonlinear mapping relationships from large-scale datasets and exhibit greater robustness and adaptability \cite{kim2022novel,vazquez2022automatic}. As machine learning continues to expand across various domains, its application in carbon emission analysis has also gained increasing traction. For instance, various studies have applied machine learning models, such as random forest, decision trees, and XGBoost \cite{qin2022estimation,shan2025unveiling}, to identify the key drivers of carbon emissions from large sets of candidate variables, often ranging from dozens to hundreds. These approaches underscore the advantages of machine learning over traditional linear models, particularly in managing high-dimensional, complex systems and capturing nonlinear relationships and interaction effects among variables \cite{qin2022estimation,ahmed2022influencing,gao2023approach}. To address the ``black-box" nature of these machine learning algorithms, some studies have further introduced approaches to improve the interpretability of results, such as SHAP (Shapley Additive Explanations) and ALE (Accumulated Local Effects). \cite{gao2023approach} and \cite{shan2025unveiling},  respectively, demonstrate how these methods can reveal the marginal contributions of variables and identify nonlinear “tipping points”, thereby enhancing the interpretability of model outputs and their value for policy guidance.

In summary, the impact of green finance on carbon emissions is complex, involving multiple interacting factors \cite{cai2025how}. However, existing literature largely overlooks the potential nonlinear relationship between green finance and carbon emissions \cite{guo2024dataset} and has yet to systematically investigate the interaction effects between green finance and other variables. These two areas remain underexplored and thus motivate our contribution.

\section{Data}
\label{sec:headings}
This research constructs a panel dataset covering 30 Chinese provinces over the period 2000–2022. All variables are measured at the provincial level, which allows us to examine how provincial characteristics influence carbon emission outcomes. Due to data limitations, Tibet, Hong Kong, Macau, and Taiwan are excluded. The following subsections provide details on the variables, their sources, and preprocessing methods.

\subsection{Dependent variable}

This study employs carbon emission intensity (CEI) as the core indicator to measure regional carbon emission performance. CEI is defined as the amount of carbon emissions per unit of Gross Domestic Product (GDP), reflecting the carbon efficiency of regional economic activities.

Compared to total carbon emissions, CEI controls differences in regional economic scale, providing a more accurate measure of green development performance across regions. This makes it more suitable for analyzing the relationship between green finance development and regional carbon emissions, as it better reflects the environmental efficiency of economic activities under varying economic contexts.

Following the method proposed by \cite{wang2022digital}, we adopt the natural logarithm of carbon emissions per unit of GDP as the response variable. The specific formula is as follows:
\[\mathrm{CEI}_{it}=\ln\frac{(\mathrm{CO}_{2})_{it}}{\mathrm{GDP}_{it}}.\]
where, for province $i$ in year $t$, CEI represents the logarithmic value of carbon emission intensity, $\mathrm{CO}_{2}$ denotes total carbon emissions, and GDP denotes gross domestic product.

\subsection{Core independent variable}

Referring to \cite{LEE2022105863}, \cite{ran2023driving} and \cite{wu2024impact}, we construct a composite green finance index as the core independent variable to measure the level of green finance development across regions. The index is based on seven secondary indicators: green bond, green fund, green credit, green insurance, green equity, green support, and green investment. These sub-indicators are first standardized to ensure comparability across provinces and then aggregated using the entropy value method of \cite{ding2016} to generate the final composite index. The entropy value method is employed to determine the weights of the sub-indicators objectively. In computing the weights for the sub-indicators, the sign (positive/negative) of each indicator signal is defined following \cite{wu2024impact}; the detailed calculation procedure is provided in \cite{LEE2022105863}. The entropy value method reduces potential bias and ensures that the resulting green finance index more accurately reflects the underlying data characteristics. A higher index value indicates a higher level of green finance development in the corresponding region. Table \ref{tab:table1} provides additional information regarding these secondary indicators.

\begin{table}[htbp]
  \caption{Description of the core independent variables}
  \centering
  \renewcommand{\arraystretch}{1.3}
  \begin{tabular}{l l l p{8cm}}
    \toprule
    Primary indicators & Secondary indicators & Weight & Variable declaration \\
    \midrule
    \multirow[c]{7}{*}{Green Financial Index (GFI)}
    & Green bond       & 0.147 & Total Green Bond Issuance / All Bond Issuance \\
    & Green fund       & 0.144 & Market Value of Green Funds / Total Market Value of All Funds \\
    & Green credit     & 0.130 & Credit for Environmental Protection Projects / Total Provincial Credit \\
    & Green insurance  & 0.129 & Revenue from Environmental Liability Insurance / Total Insurance Premium Income \\
    & Green equity     & 0.148 & Trading Volume of Carbon Emissions, Energy Use Rights, and Pollution Discharge Rights / Total Volume of Equity Market Transactions \\
    & Green support    & 0.162 & Fiscal Expenditure on Environmental Protection / General Public Budget Expenditure \\
    & Green investment & 0.140 & Environmental Pollution Control Investment / GDP \\
    \bottomrule
  \end{tabular}
  \label{tab:table1}
\end{table}

\subsection{Other independent variables}

\subsubsection{Total energy consumption}

Total Energy Consumption (TEC) is an important factor affecting CEI. Higher TEC generally increases total carbon emissions (the numerator of CEI), while greater energy use can also stimulate economic activity and expand GDP (the denominator), which may partially offset its impact on CEI. Including TEC as an explanatory variable allows us to examine its overall effect on CEI and explore its potential interaction with green finance in promoting a low-carbon transition.

\subsubsection{Climate physical risk index}
A higher climate physical risk may increase regional economic vulnerability to extreme weather, potentially affecting energy use, industrial output, and investment patterns, which in turn can influence CEI. In this context, drawing on the methodology proposed by \cite{guo2024impact}, this study constructs a Climate Physical Risk Index (CPRI) based on region-specific meteorological observations and incorporates it as one of the explanatory variables in the model. To capture both the intensity and frequency of extreme weather events, the CPRI evaluates physical climate risk across four dimensions: extreme low temperature, extreme high temperature, extreme rainfall, and extreme drought. Based on daily meteorological data from local stations, the annual number of days exceeding each threshold is calculated for every province. In order to maintain comparability across provinces, the four sub-indices are rescaled to the same interval and then aggregated into a single value indicator. As a result, a higher CPRI value indicates greater exposure to physical climate risk in a given region.

\subsubsection{Socioeconomic and environmental factors}

In addition, carbon emissions are also influenced by a range of socioeconomic and environmental factors. Building on the existing literature from Section \ref{sec:literature review}, this study incorporates several variables into the model to better capture the potential relationships among key predictors. These include the level of economic development, industrial structure, research and development (R\&D) intensity, urbanization level, population density, foreign direct investment, degree of government intervention, environmental regulation intensity. 

In light of empirical studies that have confirmed the significant regional heterogeneity in the impact of green finance on carbon emissions, this study further divides the provinces into eastern, central, and western regions based on the traditional regional classification method, and incorporates this classification as a categorical variable in the model.

\subsection{Data sources}

All data are obtained from official sources, including the China Statistical Yearbook, China Energy Statistical Yearbook, Regional Statistical Yearbooks, CSMAR database, Wind database, the Environmental Status Bulletin, and the China Deep Data database. Table \ref{tab:table2} provides a detailed description of all variables and sources.

\begin{table}[htbp]
  \caption{Data description}
  \centering
  \renewcommand{\arraystretch}{1.35}
  \begin{tabular}{l l p{6cm} p{4cm}}
    \toprule
    Index & Sign & Description & Data source \\
    \midrule
    Carbon emission intensity & CEI & Total Carbon emission / GDP & China Energy Statistical Yearbook; Regional Statistical Yearbook \\
    
    Green finance index & GFI & The index system is constructed from the seven dimensions of green credit, green equity, green insurance, green support, green investment, green bond, and green fund, and the weight is given by the entropy method & China Energy Statistical Yearbook; Regional Statistical Yearbook; Environmental Status Bulletin; Sectoral Yearbooks by Government Departments; CSMAR database \\
    
    Total energy consumption & TEC & Final energy consumption + energy transformation losses + energy losses & China Statistical Yearbook; China Energy Statistical Yearbook; Regional Statistical Yearbook \\
    
    Economic development & GDP & Per-capita GDP & China Statistical Yearbook; China Energy Statistical Yearbook; Regional Statistical Yearbook \\
    
    Foreign direct investment level & FDI & Foreign direct investment / GDP & China Statistical Yearbook; China Energy Statistical Yearbook; Regional Statistical Yearbook \\
    
    Industrial structure & industry & Output value of the tertiary sector / output value of the secondary sector & China Statistical Yearbook; China Energy Statistical Yearbook; Regional Statistical Yearbook \\
    
    Urbanization & urbanize & Urban population / total population & China Statistical Yearbook; China Energy Statistical Yearbook; Regional Statistical Yearbook \\
    
    Population density & population & Total regional population / administrative area size & China Statistical Yearbook; China Energy Statistical Yearbook; Regional Statistical Yearbook \\
    
    R\&D intensity & RDI & Internal R\&D expenditure / regional GDP & China Statistical Yearbook; China Energy Statistical Yearbook; Regional Statistical Yearbook \\
    
    Government intervention level & govern & Fiscal expenditure / regional GDP & China Statistical Yearbook; China Energy Statistical Yearbook; Regional Statistical Yearbook \\
    
    Environmental regulation & environment & Completed investment in industrial pollution control / industrial value added & China Statistical Yearbook; China Energy Statistical Yearbook; Regional Statistical Yearbook \\
    
    Climate Physical Risk Index & CPRI & Combine the four standardized sub-indices: extreme low temperature, extreme high temperature, extreme rainfall, and extreme drought using weighted average & Deep Data database \\
    \bottomrule
  \end{tabular}
  \label{tab:table2}
\end{table}

\subsection{Data preprocessing}

To reduce right-skewness in the distribution of variables, we applied logarithmic transformations to the dependent variable (CEI) and to the selected independent variables, namely economic development (GDP), population density (population), industrial structure (industry), and environmental regulation (environment), all of which exhibited a skewness value greater than 2. We then employed Generalized Variance Inflation Factors (GVIF) to conduct a preliminary analysis of potential multicollinearity among explanatory variables. A GVIF value close to 1 indicates little to no multicollinearity, while values equal to or greater than 5 or 10 suggest strong multicollinearity \cite{obrien2007caution}. Similarly, we consider the adjusted generalized standard error inflation factor (aGSIF) for which values exceeding 2.2 or 3.2 are also considered indicative of significant multicollinearity \cite{shabangu2025monthly}.

As shown in Table \ref{tab:table3}, the GVIF for the variable representing the level of economic development is 5.55 ($>$5), and the corresponding aGSIF is 2.36 ($>$2.2), indicating notable multicollinearity. Therefore, we have decided to exclude this variable from the final model specification to avoid estimation bias. 

\begin{table}[H]
  \caption{Multicollinearity results}
  \centering
  \renewcommand{\arraystretch}{1.2}
  \begin{tabular}{l c c}
    \toprule
    Variables & GVIF & aGSIF \\
    \midrule
    TEC             & 1.73 & 1.31 \\
    ln(GDP)         & \textbf{5.55} & \textbf{2.36} \\
    ln(population)  & 4.02 & 2.01 \\
    ln(industry)    & 2.24 & 1.49 \\
    ln(environment) & 1.50 & 1.22 \\
    GFI             & 2.43 & 1.56 \\
    RDI             & 4.08 & 2.03 \\
    CPRI            & 1.23 & 1.11 \\
    FDI             & 1.91 & 1.38 \\
    govern          & 2.62 & 1.62 \\
    urbanize        & 3.63 & 1.91 \\
    Region (df=2)   & 5.34 & 1.52 \\
    \bottomrule
  \end{tabular}
  \label{tab:table3}
\end{table}

\section{Methodology}
	
\subsection{Model specification}

Originally proposed by \cite{chipman2010bart}, Bayesian Additive Regression Trees (BART) is a flexible, nonparametric Bayesian ensemble method that represents the response variable $y$ as the sum of $m$ regression trees:

\[
y = \sum_{j=1}^{m} g(x; T_i, M_i) + \varepsilon, \quad \varepsilon \sim N(0, \sigma^2),
\]

where, $T_i$ denotes the structure of the $i-th$ tree, and $M_i$ represents its associated terminal node parameters. Each tree is constrained to be a weak learner, capturing only a small portion of the signal. To regularize the model and prevent overfitting, informative priors are placed on both the tree structures and terminal node values. These priors encourage shallow trees and shrinkage toward zero, thereby improving generalization and stabilizing the overall model.

Model estimation is carried out using a Bayesian backfitting Markov Chain Monte Carlo (MCMC) algorithm, implemented in the form of Gibbs sampler with Metropolis-Hastings updates. In each iteration, the sampler sequentially updates each tree $(T_i,M_i)$, conditional on the remaining $m-1$ trees. This is achieved by computing the partial residual:

\[
R_i = y - \sum_{k \neq i} g(x; T_k, M_k),
\]

which represents the part of the response not yet explained by the other trees. The residual $R_i$ serves as the local target for updating $T_i$ and $M_i$. By fitting each tree to the current residuals, the algorithm gradually refines the overall model, effectively adapting to nonlinear patterns and latent interaction effects in the data.

The update of the tree structure $T_i$ is conducted using a Metropolis–Hastings step. New tree proposals are generated through operations such as growing or pruning nodes, modifying split rules, or swapping decision nodes. The acceptance probability is computed by integrating out the terminal node parameters $M_i$, which is tractable due to the use of conjugate priors. Once the new tree is accepted, $M_i$ is updated by drawing from its full conditional posterior distribution, which follows a normal distribution. The noise variance $\sigma^2$ is updated at the end of each iteration by drawing from a full conditional inverse-gamma distribution. The MCMC algorithm was run for a large number of iterations, discarding the initial samples as burn-in. Satisfactory convergence was checked using trace plots and standard convergence diagnostics. From the resulting posterior samples, a sequence of fitted functions is obtained:

\[
f^{(k)}(x) = \sum_{i=1}^{m} g(x; T_i^{(k)}, M_i^{(k)}).
\]

These samples are treated as draws from the posterior distribution $p(f(x)|y)$, forming the basis for Bayesian inference. They can be used for prediction, uncertainty quantification, partial dependence analysis, and variable selection. The model framework and sampling algorithm together enable BART to flexibly capture complex nonlinear relationships and high-order interactions, even when such interactions are not explicitly specified in the model.

The model fit is evaluated using Root Mean Square Error (RMSE) and pseudo-$R^2$. A lower RMSE indicates a better fit to the data. The RMSE is calculated as follows:

\[
RMSE = \sqrt{\frac{1}{n} \sum_{i=1}^{n} (y_i - \hat{y}_i)^2},
\]

whereas the pseudo-$R^2$  is calculated using the formula:

\[
\text{pseudo-}R^2 = 1 - \frac{SSE}{SST}.
\]

Here, SSE is the sum of squared errors and SST is the total sum of squares, indicating the proportion of variation in the dependent variable that is explained by the model. A higher pseudo-$R^2$ suggests better explanatory power.

\subsection{Summary tools}

Despite the strong predictive performance of the BART model in both regression and classification tasks, As \cite{chipman2010bart} mentioned,  it is often considered a ``black-box" predictor due to its complex model structure which cannot be easily summarized. \cite{kapelner2016bartmachine} incorporated visualization tools such as partial dependence plots (PDPs) into the \texttt{bartMachine}  framework. However, while PDPs provide insight into average marginal effects, they may fall short in capturing more complex or localized feature interactions.  

To complement PDPs and provide a more nuanced interpretation of model behavior, this study also employs SHapley Additive exPlanations \cite{lundberg2017unified}. These tools examine feature contributions from different perspectives, thereby enhancing the depth and reliability of the model interpretation, especially for complex interactions and heterogeneous effects.

\subsubsection{Partial dependence plot}

We adopt PDPs from the \texttt{bartMachine} framework to examine how a specific predictor affects the response variable on average, while controlling for other predictors. The PDP is based on the formulation introduced by \cite{friedman2001greedy}:

\[
f_j(x_j) = E_{x_{-j}}[f(x_j, x_{-j})] := \int f(x_j, x_{-j}) dP(x_{-j}),
\]

where $f$ represents the true but unknown model function, and $dP(x_{-j})$ denotes the marginal distribution of all covariates except the $j-th$ one. The function $f_j$ measures the marginal or average effect of a single predictor $x_j$ on the response variable. Its principle is to “average out” the influence of the other variables by integrating over their marginal distribution. In other words, $f_j(x_j)$ describes how the expected model prediction changes as $x_j$ varies, while the effects of the other variables are held at their average levels, thereby isolating the independent contribution of a single predictor to the model’s output.

Since the true model function $f$ and the marginal distribution $dP(x_{-j})$  are unknown, this expectation is approximated by averaging over the training data:

\[
\hat{f}_j(x_j) = \frac{1}{n} \sum_{i=1}^{n} \hat{f}(x_j, x_{-j,i}).
\]

where $x_{-j,i}$ represents all variables except $x_j$ from the $i-th$ training sample, and $\hat{f}$ denotes predictions generated by the \texttt{bartMachine} model.

Since BART is a Bayesian method, it provides a full posterior distribution for the predictions. As a result, we can construct credible intervals for the PDP estimates. These credible bands are derived by replacing $\hat{f}$ in the above formula with functions that compute specific quantiles from the post-burn-in MCMC samples of the predicted values $\hat{y}$. This enhances both the robustness and interpretability of the PDP results.

\subsubsection{SHapley Additive exPlanations}

SHAP (SHapley Additive exPlanations) is a method grounded in the game-theoretic concept of Shapley values, which were originally proposed for model explanation by \cite{strumbelj2011general,strumbelj2014explaining}. However, the approach only gained widespread attention after \cite{lundberg2017unified} introduced SHAP as a unified framework for feature attribution. Their work not only rebranded the method but also extended its theoretical foundations, connecting Shapley values with other model-agnostic interpretability techniques such as LIME \cite{ribeiro2016why}.

SHAP treats features as ``players" in a cooperative game, where the model prediction is the ``payout" to be fairly distributed among them. The contribution of each feature is defined as its average marginal contribution across all possible coalitions of features. Formally, the Shapley value for feature $j$, given a prediction function $f$ and input $x$, is defined as:

\[
\phi_i = \sum_{S \subseteq V \setminus \{i\}} \frac{|S|! \cdot (|V| - |S| - 1)!}{|V|!} [f(S \cup i) - f(S)].
\]

where $V$ is the set of all features; $S$ is the subset of features excluding $i$; $f(S)$ is the expected prediction when only features in $S$ are known.

In practice, for models with a large number of input variables, exhaustively enumerating all feature subsets S to compute Shapley values is computationally infeasible. To address this challenge, we employed the \texttt{FastSHAP} R package developed by \cite{greenwell2020fastshap}, which provides an efficient approximation of Shapley values for any supervised learning model. Rather than calculating exact Shapley values by iterating over all possible feature coalitions, \texttt{FastSHAP} employs the Monte Carlo-based approximation method proposed by \cite{strumbelj2014explaining}. The approximated outputs, commonly referred to as SHAP values, preserve the theoretical properties of Shapley values while enabling scalable interpretation of complex black-box models.

In our analytical framework, we first use Partial Dependence Plots (PDPs) from the \texttt{bart} package to explore the global structure and interaction patterns of the model. Then, we complement PDPs with SHAP values to conduct a more in depth analysis of the interactions between variables analysis. This combined approach helps to provide a more comprehensive understanding of how input features affect the predictions, thus enhancing the transparency and reliability of BART.

\section{Results}

\subsection{Model performance}

To ensure a realistic evaluation of the model's predictive performance and to avoid data leakage, we used data from 2000 to 2020 as the training set and data from 2021 to 2022 as the test set. Additionally, we applied 10-fold cross-validation to evaluate the model's generalization ability. The BART model initially trained with default hyperparameters achieved a pseudo-R² of 0.989 and an RMSE of 0.1 in the training set. The 10-fold cross-validation yielded a pseudo-R² of 0.9586 and an RMSE of 0.1886. In the test set, the model achieved a pseudo-R² of 0.8493 and an RMSE of 0.2826. These results indicate that even with default settings, BART achieves a good fit strong fitting ability and successfully captures most of the underlying data patterns.

However, the model's predictive performance on unseen data was relatively weaker, and the residuals failed the normality test (p-value=0.00091 $<$ 0.05), indicating potential model misspecification or non-Gaussian error distribution. To address this, we used the built-in grid search function in the \texttt{bartMachine} package to tune the model's hyperparameters. The tuned model showed improved performance on the training set, with a pseudo-R² of 0.9955 and an RMSE of 0.06. Importantly, the cross-validated pseudo-R² remained stable at 0.9608, suggesting stable performance without signs of overfitting. Most notably, the pseudo-R² on the test set increased from 0.8493 to 0.8548, confirming that hyperparameter tuning enhanced the model's predictive accuracy on new data. Moreover, the residuals from the tuned model passed the normality test (p-value = 0.269 $>$ 0.05), indicating a better fit to the data.

In summary, the tuned BART model seems to achieve both high predictive power and robust generalization. Therefore, all subsequent analyses on the relationships between green finance, climate risk, and carbon emission intensity are based on the tuned model.

\subsubsection{Preliminary model diagnostics}\label{Model diagnostics and statistical significance testing}

\begin{figure}[htbp]
  \centering
  \includegraphics[width=0.47\textwidth]{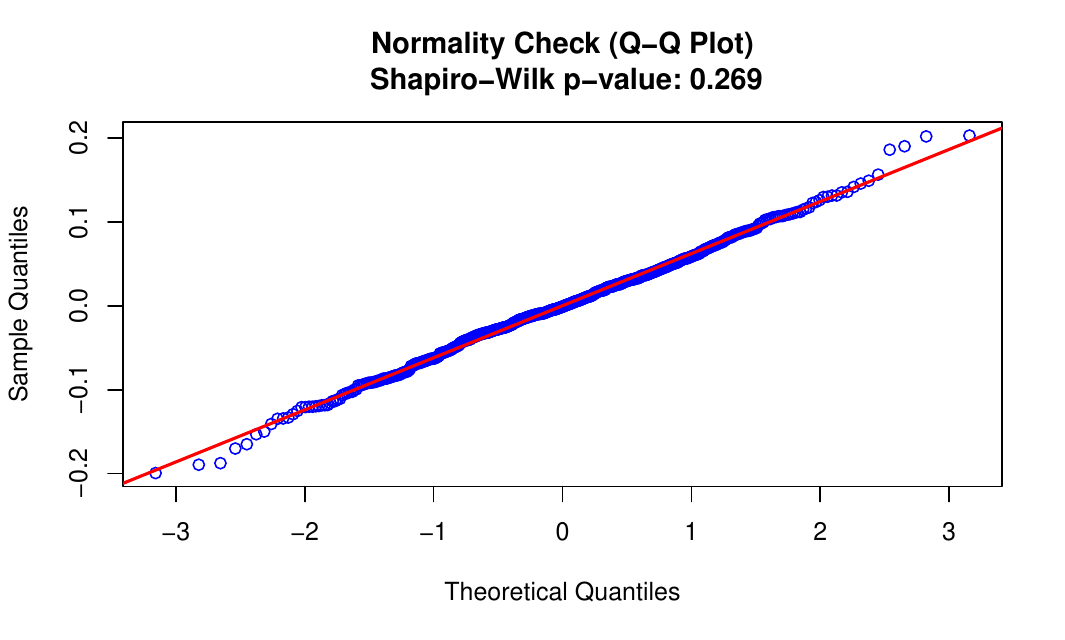}
  \includegraphics[width=0.45\textwidth]{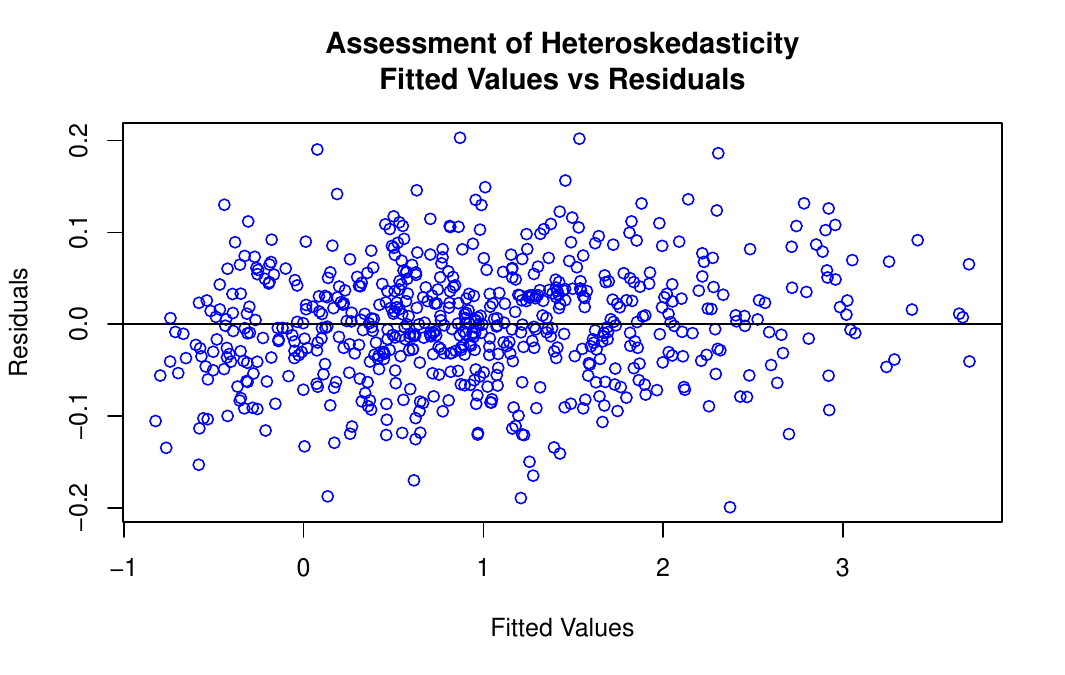}
  \caption{Test of normality of errors using QQ-plot and the Shapiro-Wilk test (left), residual plot to assess heteroskedasticity (right).}
  \label{fig:1}
\end{figure}

Figure \ref{fig:1} presents the residuals Q-Q plot to assess normality and the residuals vs predicted values plot to assess homoskedasticity. The Shapiro-Wilk test for residuals yielded a p-value of 0.269 $>$ 0.05, indicating that the assumption of normality should not be rejected. Although a few points deviate in the tails of the Q-Q plot, the majority of points align closely along the diagonal line. Meanwhile, the residuals vs predicted values plot shows no obvious pattern or trend, suggesting that there are no particular concerns regarding heteroskedasticity.

\subsection{Feature importance and interaction terms}

In addition, \texttt{bartMachine} provides a permutation-based approach to assess the statistical significance of individual predictors or groups of variables, analogous to the partial F-test in ordinary least squares regression but without assuming linearity. We applied this method to test the overall model validity, as well as the individual significance of the GFI and CPRI.

For the overall model test, we shuffled the order of the response variable $y$, thereby breaking any association between the predictors and the outcome. This is equivalent to an omnibus F-test in linear regression. The result shows a p-value of 0, well below the conventional 0.05 threshold, indicating that the model's predictive ability is highly statistically significant.

Also, we conducted individual significance tests for GFI and CPRI. In each case, the values of the target variable were randomly permuted within the sample, destroying its relationship with the response while keeping all other variables fixed. This allows us to test whether the variable contributes predictive power after controlling for the others. The results show that GFI has a p-value of 0.0396 $<$ 0.05, indicating a statistically significant effect on carbon emission intensity. In contrast, CPRI yields a p-value of 0.238, which exceeds the 0.05 threshold. Therefore, we fail to reject the null hypothesis and conclude that there is insufficient statistical evidence to support a significant effect of CPRI on carbon emission intensity.

\subsection{Feature importance and interaction terms}

This section utilizes the built-in functionalities of \texttt{bartMachine} to examine the importance of individual predictors and their interactions within BART. In the BART framework, variable importance is evaluated using the inclusion proportion metric \cite{chipman2010bart, bleich2014variable}. The inclusion proportion for a given predictor represents the frequency with which that variable is selected as a splitting rule across all posterior samples (among all tree structures generated during the MCMC iterations). A higher inclusion proportion indicates that the variable appears more frequently in the model’s partitioning process and thus contributes more substantially to explaining variations in the response variable. 

Interaction effects are evaluated following the approach of \cite{damien2013bayesian}, where an interaction between two predictors is considered present when both variables appear within splitting rules along a path in the same tree. To quantify these interactions, the method counts how often each pair of variables co-occurs in such splits across all trees and posterior samples. Summing these counts provides an overall measure of the interaction strength for each pair, which can then be used to assess the relative importance of their combined effect on the response. This approach allows the identification of predictor pairs whose joint influence is substantial, including cases where a variable may interact with itself.

\begin{figure}[htbp]
  \centering
  \includegraphics[width=0.5\textwidth]{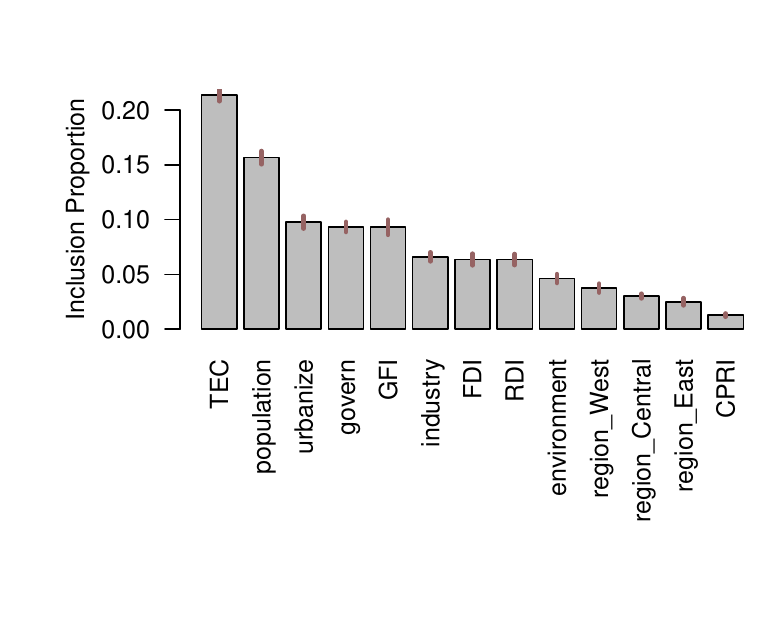}
  \caption{The average inclusion proportions of explanatory variables in the tuned BART model, reflecting how frequently each variable appears in the model's decision paths. The y-axis represents the average inclusion proportions calculated over 100 model constructions. The segments atop the bars indicate the 95\% confidence intervals.}
  \label{fig:2}
\end{figure}

As shown in Figure \ref{fig:2}, TEC appears most frequently in the model’s decision paths and has relatively narrow 95\% credible intervals, indicating that it contributes the most to explaining variations in CEI, with a high degree of certainty. GFI ranks fifth, and its 95\% credible interval is relatively wide compared with those of other predictors, suggesting that although green finance exerts a significant influence, the model exhibits some uncertainty about the precise magnitude of its effect. In contrast, CPRI ranks the lowest, with a near-zero inclusion proportion and one of the narrowest 95\% credible interval, suggesting that it plays a limited role in explaining variations in CEI and that the model is fairly certain about this result. The statistical significance of each variable are further validated in Section \ref{Model diagnostics and statistical significance testing}.

\begin{figure}[htbp]
  \centering
  \includegraphics[width=0.5\textwidth]{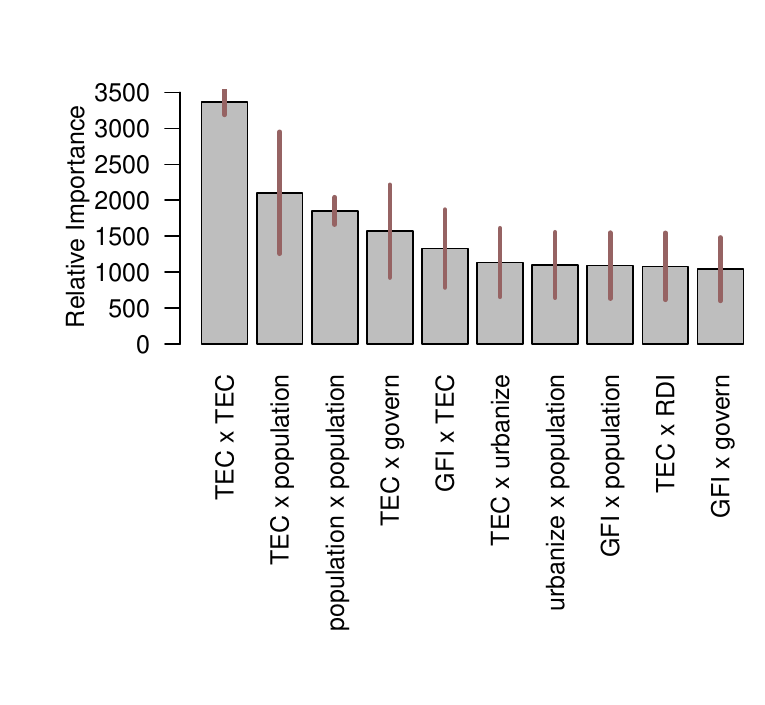}
  \caption{In the tuned BART model, this figure displays the top 10 variable pairs ranked by average interaction counts, based on 25 model constructions. The vertical axis represents the average number of interactions, and the segments atop the bars indicate 95\% confidence intervals.}
  \label{fig:3}
\end{figure}

In terms of interaction effects, \texttt{bartMachine} constructs 25 independent BART models, each trained from a different random initialization (randomly chosen initial tree structure and parameter state). The interaction strengths are then averaged across these models to obtain stable estimates that account for different posterior modes in the tree and distribution. Figure~\ref{fig:3} displays the top 10 variable pairs with the strongest interactions, as measured by their average co-occurrence across decision paths in 25 constructed models. Among these, GFI appears frequently as a component of key interacting pairs, including GFI × TEC, GFI × population, and GFI × govern. The presence of GFI in multiple high-ranking interactions indicates that its influence on CEI may be highly conditional on other structural and socioeconomic factors. In particular, the interaction between GFI and TEC ranks fifth overall and stands out as the most influential GFI-related pair. This finding suggests that the marginal impact of green finance on carbon emissions may be amplified or moderated depending on the level of energy consumption. Therefore, the subsequent analysis will focus on the interaction between GFI and TEC to explore the mechanism through which green finance influences the marginal effect of energy consumption on carbon emission intensity.

\subsection{Nonlinear and interaction effects analysis}

\begin{figure}[htbp]
  \centering
  \includegraphics[width=0.5\textwidth]{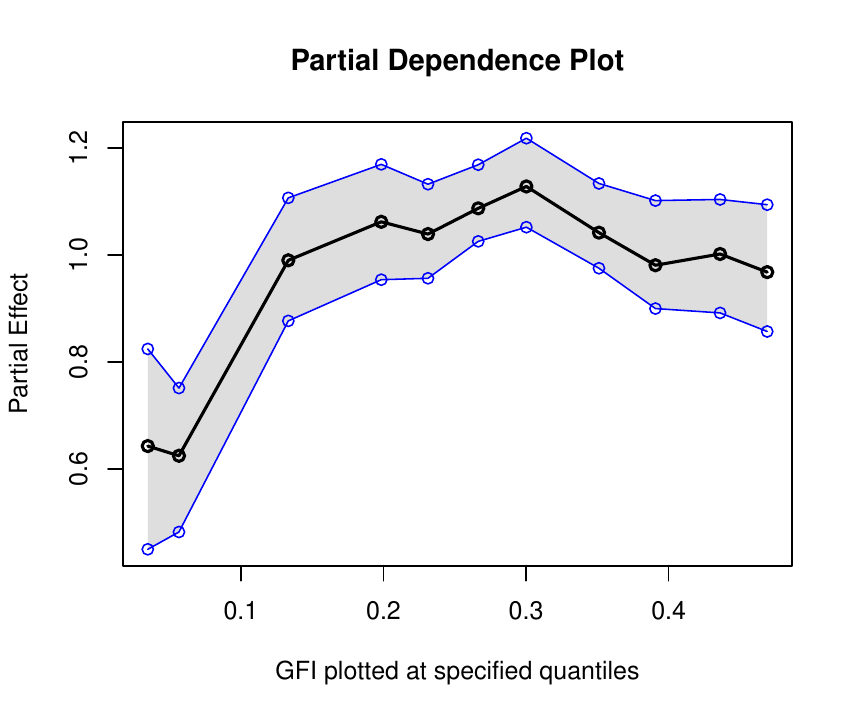}
  \caption{PDP of GFI on CEI. The black line represents the average marginal effect of GFI on CEI when all other variables are held constant. The shaded gray area and blue lines indicate the 95\% credible interval, reflecting model uncertainty. The narrowing of the credible interval suggests increasing model certainty. The x-axis shows GFI values, while the y-axis represents the average predicted CEI at each GFI level, conditional on other variables being fixed.}
  \label{fig:4}
\end{figure}

\begin{figure}[htbp]
  \centering
  \includegraphics[width=0.5\textwidth]{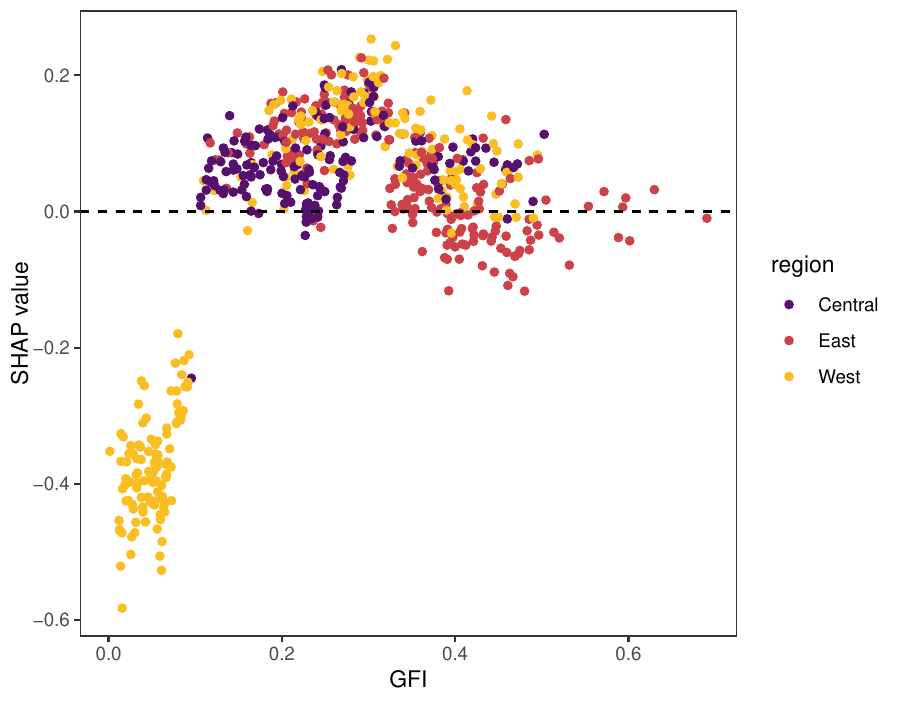}
  \caption{SHAP PDP of GFI on CEI, color-coded by region (red for Eastern, purple for Central, yellow for Western). Each point represents an observation, with the x-axis showing the value of GFI and the y-axis showing the corresponding SHAP value, which reflects the marginal contribution of GFI to the model prediction of CEI. SHAP values represent the marginal contribution of a given feature (GFI) to the model's prediction of CEI. A positive SHAP value indicates that GFI increases predicted CEI (i.e., contributes positively), while a negative value suggests a suppressing effect (i.e., contributes negatively).}
  \label{fig:5}
\end{figure}

Figures \ref{fig:4} and \ref{fig:5} jointly explore the nonlinear and heterogeneous effects of GFI on CEI, using PDP from BART and SHAP visualizations. The PDP curve in Figure \ref{fig:4} exhibits a typical inverted U-shaped pattern: at lower GFI levels, CEI increases with GFI, but once GFI exceeds a threshold of approximately 0.3, the curve begins to decline, indicating an emission-reduction effect at higher levels of GFI. However, the interpretability of this nonlinear pattern depends on the underlying data-generating process.

The PDP is constructed by averaging over the effects of other variables while allowing GFI to vary, which implicitly assumes that GFI is independent or only weakly correlated with other covariates. In real economic systems, this assumption rarely holds, particularly in the low-GFI range where green finance is often closely related to factors such as technological progress, industrial structure, and policy implementation. Therefore, the apparent “positive correlation” observed at low GFI levels is more likely to reflect unobserved confounding effects rather than the true marginal impact of GFI. For example, in regions with lower technological development or less optimized economic structures, both insufficient green finance and higher carbon emissions tend to occur simultaneously, resulting in a spurious positive relationship in the model.

Once GFI exceeds the threshold of approximately 0.3, the influence of these confounding factors gradually weakens, and the independent role of green finance begins to emerge. The downward trend in this range more likely reflects the genuine emission-reduction effect of green finance. On one hand, a higher level of GFI channels financial and policy resources toward low-carbon, energy-efficient, and renewable energy projects, directly lowering carbon emission intensity. On the other hand, the expansion of green finance may also stimulate technological innovation and green industrial upgrading, establishing a long-term mitigation mechanism. In addition, a mature green finance system can enhance the environmental governance capacity of enterprises and local governments, leading to more efficient energy use and emission control. Therefore, the most meaningful part of Figure \ref{fig:4} lies in the high-GFI range, where green finance demonstrates a tangible and economically interpretable marginal effect on reducing CEI. This indicates that the environmental benefits of green finance become effective only when it reaches a sufficiently advanced level of development.

Figure \ref{fig:5} presents the SHAP-based analysis, which further reveals the regional heterogeneity underlying the nonlinear relationship observed in the PDP results. As shown in the figure, most samples in the range of GFI below approximately 0.3 are concentrated in the central and western regions, whereas those from the eastern region are primarily distributed at higher GFI levels above 0.3. It is important to note that although the SHAP values for the western region are generally negative at low GFI levels, this does not imply that low GFI exerts a genuine emission-reduction effect. As discussed in the above, low GFI levels are often accompanied by limited technological innovation capacity, slower industrial upgrading, and weaker policy implementation. Therefore, the relationship observed in this range likely reflects unobserved structural factors or confounding effects rather than the true marginal contribution of GFI.

Beyond the threshold of approximately 0.3, the carbon reduction effect of GFI becomes evident and displays notable regional heterogeneity. In the central and western regions, SHAP values exhibit an overall downward trend but remain mostly positive, suggesting that although green finance has developed to a higher level, its carbon reduction potential has not yet been fully realized. These regions may still be in a stage dominated by investment expansion and institutional development, where economic foundations, technological capabilities, and policy frameworks require further improvement. In contrast, SHAP values in the eastern region show a more pronounced decline at higher GFI levels, with some observations even turning negative. This indicates that in regions with stronger economic foundations, greater technological innovation, and more mature policy systems, advanced green finance development has begun to exert a substantial and measurable mitigating effect on carbon emission intensity, reflecting the effectiveness of a well-established and efficient green finance system.

Taken together with the PDP results, these findings demonstrate that the carbon reduction effects of green finance are not only nonlinear but also regionally heterogeneous. More importantly, the genuine marginal impact of green finance tends to emerge only once it surpasses a certain development threshold, and both the strength and timing of this effect vary significantly across regions. This evidence underscores the need for region-specific strategies that align with local economic structures and development stages to promote green finance as a sustained mechanism for reducing carbon intensity.

\subsubsection{Interaction effects between GFI and TEC}

\begin{figure}[htbp]
  \centering
  \includegraphics[width=0.42\textwidth]{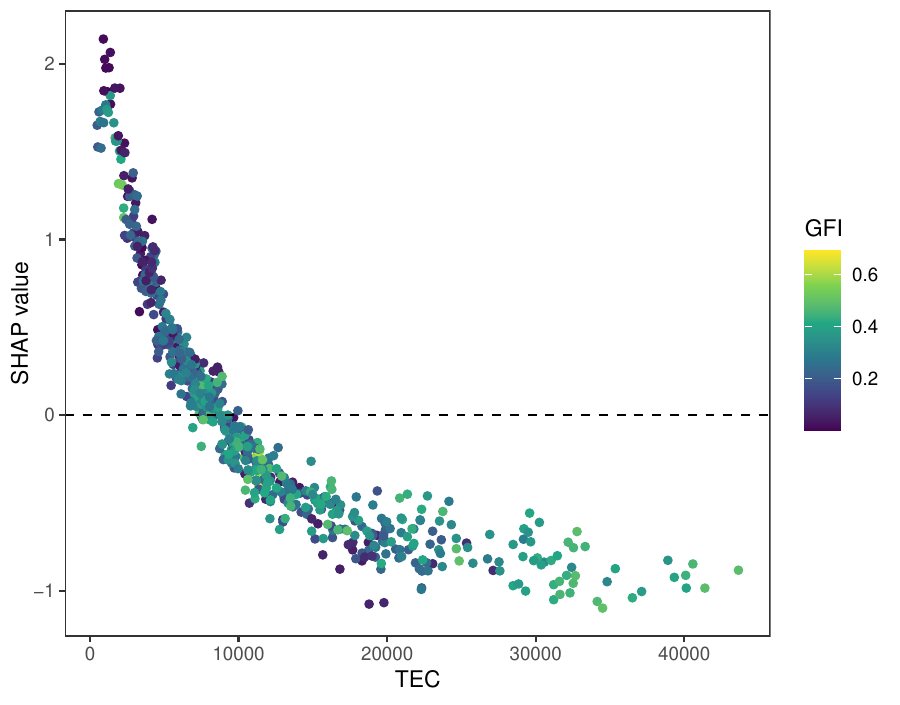}
  \caption{SHAP interaction plots between GFI and TEC. The plot shows the SHAP values of TEC across varying levels of GFI. The x-axis represents TEC, and each point is colored according to the GFI level, with lighter colors (yellow) indicating higher levels of green finance development. The y-axis represents the SHAP value of TEC, reflecting its marginal contribution to predicted CEI. This visualization highlights the combined marginal effects of the two features rather than their pure interaction term.}
  \label{fig:6}
\end{figure}

Figure \ref{fig:6} presents the SHAP interaction effects between total energy consumption (TEC) and green finance (GFI), illustrating how their joint variations influence carbon emission intensity (CEI). It can be observed that when both TEC and GFI are relatively low, the SHAP values are predominantly positive, indicating that limited energy use and underdeveloped green finance are associated with higher CEI levels. As TEC and GFI increase, the SHAP values exhibit a downward trend, eventually even turning negative. It should be noted, however, that this decline in SHAP values does not imply that higher energy consumption directly reduces carbon intensity.

This phenomenon can be explained by the definition of CEI and the potential role of green finance. In this study, CEI is defined as carbon emissions per unit of GDP, and higher TEC is generally accompanied by increased economic output, which expands the GDP denominator and leads to a lower measured CEI. High levels of green finance may promote improvements in energy efficiency and transitions to cleaner energy sources, thereby weakening the marginal impact of energy consumption on CEI. Moreover, in more advanced stages of green finance development, economic scale expansion may offset part of the emission-promoting effect of increased energy consumption, further reducing carbon intensity per unit of output. In other words, the observed decline in SHAP values reflects the combined moderating effects of green finance and economic growth on carbon intensity, rather than a direct emission-reducing effect of energy consumption.

Overall, the interaction between GFI and TEC reveals a clear nonlinear and synergistic relationship in shaping carbon emission intensity. Specifically, green finance not only directly influences CEI but also plays a critical moderating role by weakening the marginal effect of energy consumption on carbon intensity. These findings emphasize the importance of integrating financial instruments into carbon mitigation strategies. In particular, energy-intensive regions stand to benefit more from targeted green finance policies that can curb emissions even amid rising energy demand. Moreover, accounting for such interaction effects, including TEC × GFI, contributes to a more nuanced understanding of how various policy instruments, such as financial incentives, energy regulation, and regional planning, can be effectively coordinated to achieve long-term environmental objectives.

\section{Conclusion}

This study employs BART to investigate the complex relationships between green finance and carbon emissions across 30 Chinese provinces from 2000 to 2022. Through careful model tuning, evaluation and diagnostic testing, we confirm the strong predictive power and robustness of the BART approach in modeling environmental-economic interactions. The tuned model achieves high accuracy and generalizability, making it a reliable basis for further interpretation. Our analysis yields several key findings: (1) green finance is a relevant predictor of carbon emissions, but with a non-linear pattern. The impact of GFI on CEI exhibits a distinct inverted U-shaped pattern. This result generally aligns with the previous findings of \cite{wu2024impact}. At low levels of GFI, the observed positive relationship with CEI likely reflects confounding structural factors rather than a direct effect of green finance. Once GFI surpasses a threshold of approximately 0.3, its genuine emission-reduction effect emerges, highlighting the importance of advanced development and long-term commitment in green financial systems. (2) Regional heterogeneity is apparent. The effects of GFI on CEI vary across regions. Eastern provinces, where green finance has a stronger presence, show the most pronounced carbon reduction effects at higher GFI levels. Central and Western regions, in contrast, remain in a transitional stage, with the mitigation potential of green finance not yet fully realized. These differences emphasize the need for region-specific strategies that account for local economic structures, technological capabilities, and policy maturity. (3) Green finance interacts with energy consumption, jointly influencing the outcome. GFI moderates the effects of TEC on CEI. High GFI can offset the positive emissions effects typically associated with high TEC, reflecting that green finance can weaken the correlation between energy consumption and carbon emissions under certain conditions. This underscores the importance of integrating financial instruments into carbon mitigation strategies, particularly in energy-intensive regions; (4) CPRI is not statistically significant. Contrary to expectations, CPRI does not exhibit a statistically significant impact on CEI in our model. However, this finding should not be interpreted as evidence that climate risk is irrelevant to carbon emissions. Rather, it may stem from the limitations of using annual, province-level data, which may not adequately capture the more nuanced or delayed effects of climate risk. Climate risk often exerts its influence with a temporal lag and is likely to have a more pronounced impact at the micro level, particularly on individual firms or sectors. This highlights the need for future research that utilizes more granular data and incorporates potential lag structures to better assess the complex relationship between climate risk and emissions.

These findings suggest that while green finance is a powerful tool for reducing carbon emissions, its design and implementation must account for nonlinear effects, interaction dynamics and regional heterogeneity. Based on these findings, the analysis can be further extended to incorporate a spatial dimension by using geographical information across provinces to examine whether spatial spillover effects or regional clustering patterns influence the effectiveness of green finance policies and carbon reduction outcomes.
\newpage


\bibliographystyle{unsrt}


\begin{thebibliography}{99}

\bibitem{wang2023can}
J.~Wang, J.~Tian, Y.~Kang, and K.~Guo,
\newblock Can green finance development abate carbon emissions: Evidence from China,
\newblock \emph{International Review of Economics \& Finance}, vol.~88, pp.~73--91, 2023.
\newblock DOI: \url{10.1016/j.iref.2023.06.011}

\bibitem{wu2024impact}
G.~Wu, X.~Liu, and Y.~Cai,
\newblock The impact of green finance on carbon emission efficiency,
\newblock \emph{Heliyon}, vol.~10, no.~1, 2024.
\newblock DOI: \url{10.1016/j.heliyon.2023.e23803}

\bibitem{niyazbekova2021growth}
S.~Niyazbekova, B.~Jazykbayeva, A.~Mottaeva, E.~Beloussova, B.~Suleimenova, and A.~Zueva,
\newblock The growth of "green" finance at the global level in the context of sustainable economic development,
\newblock in \emph{E3S Web of Conferences}, vol.~244, pp.~10058, 2021.
\newblock DOI: \url{10.1051/e3sconf/202124410058}

\bibitem{lin2023impact}
Z.~Lin, H.~Wang, W.~Li, and M.~Chen,
\newblock Impact of green finance on carbon emissions based on a two-stage LMDI decomposition method,
\newblock \emph{Sustainability}, vol.~15, no.~17, pp.~12808, 2023.
\newblock DOI: \url{10.3390/su151712808}

\bibitem{du2025decarbonization}
M.~Du, J.~Zhang, and X.~Hou,
\newblock Decarbonization like China: How does green finance reform and innovation enhance carbon emission efficiency?,
\newblock \emph{Journal of Environmental Management}, vol.~376, pp.~124331, 2025.
\newblock DOI: \url{10.1016/j.jenvman.2025.124331}

\bibitem{Institute2023Fintech}
Institute of Finance and Sustainability and Paulson Institute,
\newblock Fintech facilitates green finance in China: Cases and outlook,
\newblock 2023. \url{https://paulsoninstitute.org.cn/wp-content/uploads/2023/10/2023-Fintech-Report_Full-Report_Final.pdf}

\bibitem{lee2024impact}
C.~C.~Lee, H.~Song, and J.~An,
\newblock The impact of green finance on energy transition: Does climate risk matter?,
\newblock \emph{Energy Economics}, vol.~129, pp.~107258, 2024.
\newblock DOI: \url{10.1016/j.eneco.2023.107258}

\bibitem{lu2025climate}
H.~Lu and X.~Wang,
\newblock Climate change news sensitivity and expected stock returns: Evidence from China,
\newblock \emph{Finance Research Letters}, pp.~107497, 2025.
\newblock DOI: \url{10.1016/j.frl.2025.107497}

\bibitem{york2003stirpat}
R.~York, E.~A.~Rosa, and T.~Dietz,
\newblock STIRPAT, IPAT and ImPACT: Analytic tools for unpacking the driving forces of environmental impacts,
\newblock \emph{Ecological Economics}, vol.~46, no.~3, pp.~351--365, 2003.
\newblock DOI: \url{10.1016/S0921-8009(03)00188-5}

\bibitem{yang2018what}
L.~Yang, H.~Xia, X.~Zhang, and S.~Yuan,
\newblock What matters for carbon emissions in regional sectors? A China study of extended STIRPAT model,
\newblock \emph{Journal of Cleaner Production}, vol.~180, pp.~595--602, 2018.
\newblock DOI: \url{10.1016/j.jclepro.2018.01.116}

\bibitem{tang2024study}
X.~Tang, S.~Liu, Y.~Wang, and others,
\newblock Study on carbon emission reduction countermeasures based on carbon emission influencing factors and trends,
\newblock \emph{Environmental Science and Pollution Research}, vol.~31, pp.~14003--14022, 2024.
\newblock DOI: \url{10.1007/s11356-024-31962-6}

\bibitem{xiao2023decomposition}
M.~Xiao and X.~Peng,
\newblock Decomposition of carbon emission influencing factors and research on emission reduction performance of energy consumption in China,
\newblock \emph{Frontiers in Environmental Science}, vol.~10, pp.~1096650, 2023.
\newblock DOI: \url{10.3389/fenvs.2022.1096650}

\bibitem{chen2021analysis}
J.~Chen, X.~Lian, H.~Su, and others,
\newblock Analysis of China’s carbon emission driving factors based on the perspective of eight major economic regions,
\newblock \emph{Environmental Science and Pollution Research}, vol.~28, pp.~8181--8204, 2021.
\newblock DOI: \url{10.1007/s11356-020-11044-z}

\bibitem{li2022dynamic}
Z.~Li and J.~Wang,
\newblock The dynamic impact of digital economy on carbon emission reduction: Evidence city-level empirical data in China,
\newblock \emph{Journal of Cleaner Production}, vol.~351, pp.~131570, 2022.
\newblock DOI: \url{10.1016/j.jclepro.2022.131570}

\bibitem{chen2019driving}
J.~Chen, C.~Xu, L.~Cui, S.~Huang, and M.~Song,
\newblock Driving factors of CO\textsubscript{2} emissions and inequality characteristics in China: A combined decomposition approach,
\newblock \emph{Energy Economics}, vol.~78, pp.~589--597, 2019.
\newblock DOI: \url{10.1016/j.eneco.2018.12.011}

\bibitem{liu2019driving}
B.~Liu, J.~Shi, H.~Wang, X.~Su, and P.~Zhou,
\newblock Driving factors of carbon emissions in China: A joint decomposition approach based on meta-frontier,
\newblock \emph{Applied Energy}, vol.~256, pp.~113986, 2019.
\newblock DOI: \url{10.1016/j.apenergy.2019.113986}

\bibitem{zhang2022carbon}
L.~Zhang, Y.~Yan, W.~Xu, J.~Sun, and Y.~Zhang,
\newblock Carbon emission calculation and influencing factor analysis based on industrial big data in the "double carbon" era,
\newblock \emph{Computational Intelligence and Neuroscience}, vol.~2022, no.~1, pp.~2815940, 2022.
\newblock DOI: \url{10.1155/2022/2815940}

\bibitem{sun2022characteristics}
Z.~Y.~Sun, M.~X.~Deng, D.~Li, and Y.~Sun,
\newblock Characteristics and driving factors of carbon emissions in China,
\newblock \emph{Journal of Environmental Planning and Management}, vol.~67, no.~5, pp.~967--992, 2022.
\newblock DOI: \url{10.1080/09640568.2022.2142906}

\bibitem{lin2023china}
Z.~Lin, X.~Liao, and Y.~Yang,
\newblock China’s experience in developing green finance to reduce carbon emissions: From spatial econometric model evidence,
\newblock \emph{Environmental Science and Pollution Research}, vol.~30, pp.~15531--15547, 2023.
\newblock DOI: \url{10.1007/s11356-022-23246-8}

\bibitem{wang2022how}
J.~Wang and Y.~Ma,
\newblock How does green finance affect CO\textsubscript{2} emissions? Heterogeneous and mediation effects analysis,
\newblock \emph{Frontiers in Environmental Science}, vol.~10, pp.~931086, 2022.
\newblock DOI: \url{10.3389/fenvs.2022.931086}

\bibitem{tu2025impact}
W.~Tu, Q.~Ma, X.~Zhao, and W.~Liu,
\newblock The impact of green finance on carbon emissions based on fixed effects model,
\newblock \emph{IEEE Access}, 2025.
\newblock DOI: \url{10.1109/ACCESS.2025.3534240}

\bibitem{ran2023driving}
C.~Ran and Y.~Zhang,
\newblock The driving force of carbon emissions reduction in China: Does green finance work,
\newblock \emph{Journal of Cleaner Production}, vol.~421, pp.~138502, 2023.
\newblock DOI: \url{10.1016/j.jclepro.2023.138502}

\bibitem{zhao2024role}
X.~Zhao and X.~Li,
\newblock The role of green finance in mitigating climate change risks: A quantitative analysis of sustainable investments,
\newblock \emph{Environmental Science and Pollution Research}, vol.~31, pp.~7569--7585, 2024.
\newblock DOI: \url{10.1007/s11356-023-31705-z}

\bibitem{ma2024research}
Z.~Ma and Z.~Fei,
\newblock Research on the mechanism of the carbon emission reduction effect of green finance,
\newblock \emph{Sustainability}, vol.~16, no.~7, pp.~3087, 2024.
\newblock DOI: \url{10.3390/su16073087}

\bibitem{yu2024machine}
W.~Yu, L.~Xia, and Q.~Cao,
\newblock A machine learning algorithm to explore the drivers of carbon emissions in Chinese cities,
\newblock \emph{Scientific Reports}, vol.~14, pp.~23609, 2024.
\newblock DOI: \url{10.1038/s41598-024-75753-y}

\bibitem{cai2025how}
Q.~Cai, W.~Chen, M.~Wang, and K.~Di,
\newblock How does green finance influence carbon emission intensity? A non-linear fsQCA-ANN approach,
\newblock \emph{Polish Journal of Environmental Studies}, vol.~34, no.~5, 2025.
\newblock DOI: \url{10.15244/pjoes/190658}

\bibitem{li2017industrial}
W.~Li, W.~Wang, Y.~Wang, and others,
\newblock Industrial structure, technological progress and CO\textsubscript{2} emissions in China: Analysis based on the STIRPAT framework,
\newblock \emph{Natural Hazards}, vol.~88, pp.~1545--1564, 2017.
\newblock DOI: \url{10.1007/s11069-017-2932-1}

\bibitem{zhang2019identifying}
S.~Zhang and T.~Zhao,
\newblock Identifying major influencing factors of CO\textsubscript{2} emissions in China: Regional disparities analysis based on STIRPAT model from 1996 to 2015,
\newblock \emph{Atmospheric Environment}, vol.~207, pp.~136--147, 2019.
\newblock DOI: \url{10.1016/j.atmosenv.2018.12.040}

\bibitem{thio2022estimation}
E.~Thio, M.~Tan, L.~Li, and others,
\newblock The estimation of influencing factors for carbon emissions based on EKC hypothesis and STIRPAT model: Evidence from top 10 countries,
\newblock \emph{Environment, Development and Sustainability}, vol.~24, pp.~11226--11259, 2022.
\newblock DOI: \url{10.1007/s10668-021-01905-z}

\bibitem{kim2022novel}
J.~Kim, J.~Yu, C.~Kang, G.~Ryang, Y.~Wei, and X.~Wang,
\newblock A novel hybrid water quality forecast model based on real-time data decomposition and error correction,
\newblock \emph{Process Safety and Environmental Protection}, vol.~162, pp.~553--565, 2022.
\newblock DOI: \url{10.1016/j.psep.2022.04.020}

\bibitem{vazquez2022automatic}
D.~Vazquez, R.~Guimera, M.~Sales-Pardo, and G.~Guillen-Gosalbez,
\newblock Automatic modeling of socioeconomic drivers of energy consumption and pollution using Bayesian symbolic regression,
\newblock \emph{Sustainable Production and Consumption}, vol.~30, pp.~596--607, 2022.
\newblock DOI: \url{10.1016/j.spc.2021.12.025}

\bibitem{qin2022estimation}
J.~Qin and N.~Gong,
\newblock The estimation of the carbon dioxide emission and driving factors in China based on machine learning methods,
\newblock \emph{Sustainable Production and Consumption}, vol.~33, pp.~218--229, 2022.
\newblock DOI: \url{10.1016/j.spc.2022.06.027}

\bibitem{ahmed2022influencing}
M.~Ahmed, C.~Shuai, and M.~Ahmed,
\newblock Influencing factors of carbon emissions and their trends in China and India: A machine learning method,
\newblock \emph{Environmental Science and Pollution Research}, vol.~29, pp.~48424--48437, 2022.
\newblock DOI: \url{10.1007/s11356-022-18711-3}

\bibitem{gao2023approach}
P.~Gao, C.~Zhu, Y.~Zhang, and B.~Chen,
\newblock An approach for analyzing urban carbon emissions using machine learning models,
\newblock \emph{Indoor and Built Environment}, vol.~32, no.~8, pp.~1657--1667, 2023.
\newblock DOI: \url{10.1177/1420326X231162253}

\bibitem{shan2025unveiling}
T.~Shan, S.~Feng, K.~Li, R.~Chang, and R.~Huang,
\newblock Unveiling the effects of artificial intelligence and green technology convergence on carbon emissions: An explainable machine learning-based approach,
\newblock \emph{Journal of Environmental Management}, vol.~373, pp.~123657, 2025.
\newblock DOI: \url{10.1016/j.jenvman.2024.123657}

\bibitem{guo2024impact}
X.~Guo, J.~Yang, Y.~Shen, and X.~Zhang,
\newblock Impact on green finance and environmental regulation on carbon emissions: Evidence from China,
\newblock \emph{Frontiers in Environmental Science}, vol.~12, pp.~1307313, 2024.
\newblock DOI: \url{10.3389/fenvs.2024.1307313}

\bibitem{wang2022digital}
Q.~Wang, A.~Hu, and Z.~Tian,
\newblock Digital transformation and electricity consumption: Evidence from the Broadband China pilot policy,
\newblock \emph{Energy Economics}, vol.~115, pp.~106346, 2022.
\newblock DOI: \url{10.1016/j.eneco.2022.106346}

\bibitem{LEE2022105863}
C.~C. Lee and C.~C. Lee,
\newblock How does green finance affect green total factor productivity? Evidence from China,
\newblock \emph{Energy Economics}, vol.~107, pp.~105863, 2022.
\newblock DOI: \url{https://doi.org/10.1016/j.eneco.2022.105863}

\bibitem{ding2016}
L.~Ding, Z.~Shao, H.~Zhang, C.~Xu, and D.~Wu,
\newblock A comprehensive evaluation of urban sustainable development in China based on the TOPSIS-Entropy method,
\newblock \emph{Sustainability}, vol.~8, no.~8, pp.~746, 2016.
\newblock DOI: \url{10.3390/su8080746}

\bibitem{guo2024dataset}
K.~Guo, Q.~Ji, and D.~Zhang,
\newblock A dataset to measure global climate physical risk,
\newblock \emph{Data in Brief}, vol.~54, pp.~110502, 2024.
\newblock DOI: \url{10.1016/j.dib.2024.110502}

\bibitem{obrien2007caution}
R.~M. O'Brien,
\newblock A caution regarding rules of thumb for variance inflation factors,
\newblock \emph{Quality \& Quantity}, vol.~41, pp.~673--690, 2007.
\newblock DOI: \url{10.1007/s11135-006-9018-6}

\bibitem{shabangu2025monthly}
F.~W. Shabangu, K.~Hlati, M.~A. van~den~Berg, T.~Lamont, and S.~P. Kirkman,
\newblock Monthly and diel acoustic occurrence of four baleen whale species in South African waters,
\newblock \emph{Ecology and Evolution}, vol.~15, no.~8, pp.~e72004, 2025.
\newblock DOI: \url{10.1002/ece3.72004}

\bibitem{chipman2010bart}
H.~A. Chipman, E.~I. George, and R.~E. McCulloch,
\newblock BART: Bayesian additive regression trees,
\newblock \emph{The Annals of Applied Statistics}, vol.~4, no.~1, pp.~266--298, 2010.
\newblock DOI: \url{10.1214/09-AOAS285}

\bibitem{kapelner2016bartmachine}
A.~Kapelner and J.~Bleich,
\newblock bartMachine: Machine learning with Bayesian additive regression trees,
\newblock \emph{Journal of Statistical Software}, vol.~70, no.~4, pp.~1--40, 2016.
\newblock DOI: \url{10.18637/jss.v070.i04}

\bibitem{friedman2001greedy}
J.~H. Friedman,
\newblock Greedy function approximation: A gradient boosting machine,
\newblock \emph{Annals of Statistics}, pp.~1189--1232, 2001.
\newblock URL: \url{https://www.jstor.org/stable/2699986}

\bibitem{strumbelj2011general}
E.~\v{S}trumbelj and I.~Kononenko,
\newblock A general method for visualizing and explaining black-box regression models,
\newblock in \emph{Adaptive and Natural Computing Algorithms}, Lecture Notes in Computer Science, vol.~6594, pp.~21--30, 2011.
\newblock Springer, Berlin, Heidelberg.
\newblock DOI: \url{10.1007/978-3-642-20267-4_3}

\bibitem{strumbelj2014explaining}
E.~\v{S}trumbelj and I.~Kononenko,
\newblock Explaining prediction models and individual predictions with feature contributions,
\newblock \emph{Knowledge and Information Systems}, vol.~41, pp.~647--665, 2014.
\newblock DOI: \url{10.1007/s10115-013-0679-x}

\bibitem{lundberg2017unified}
S.~M. Lundberg and S.~I. Lee,
\newblock A unified approach to interpreting model predictions,
\newblock in \emph{Advances in Neural Information Processing Systems}, vol.~30, 2017.
\newblock URL: \url{https://proceedings.neurips.cc/paper_files/paper/2017/file/8a20a8621978632d76c43dfd28b67767-Paper.pdf}

\bibitem{ribeiro2016why}
M.~T. Ribeiro, S.~Singh, and C.~Guestrin,
\newblock ``Why should I trust you?'': Explaining the predictions of any classifier,
\newblock in \emph{Proceedings of the 22nd ACM SIGKDD International Conference on Knowledge Discovery and Data Mining},
pp.~1135--1144, 2016.
\newblock DOI: \url{10.1145/2939672.2939778}

\bibitem{greenwell2020fastshap}
B.~Greenwell and M.~B. Greenwell,
\newblock Package \texttt{fastshap},
\newblock 2020.
\newblock URL: \url{https://cran.r-project.org/web/packages/fastshap/fastshap.pdf}

\bibitem{bleich2014variable}
J.~Bleich, A.~Kapelner, E.~I. George, and S.~T. Jensen,
\newblock Variable selection for BART: an application to gene regulation,
\newblock \emph{The Annals of Applied Statistics}, vol.~8, no.~3, pp.~1750--1781, 2014.
\newblock DOI: \url{10.1214/14-AOAS755}

\bibitem{damien2013bayesian}
P.~Damien, P.~Dellaportas, N.~Polson, and D.~Stephens,
\newblock \emph{Bayesian Theory and Applications},
\newblock 1st ed., Oxford University Press, 2013.



\end{thebibliography}


\end{document}